\documentclass{moriond}

\usepackage{subcaption}
\usepackage{amsmath,amssymb}
\usepackage[font=scriptsize]{caption}

\def\be{\begin{equation}}
\def\ee{\end{equation}}
\def\bea{\begin{eqnarray}}
\def\eea{\end{eqnarray}}

\newcommand{\Photo}{}

\begin{document}
\vspace*{4cm}
\title{Toward the first cosmological results of the NIKA2 Sunyaev-Zeldovich Large Program:
The SZ-Mass scaling relation
}
\author{ A.~Moyer-Anin$^1$, 
R.~Adam$^2$, 
P.~Ade$^3$, 
H.~Ajeddig$^4$, 
P.~Andr\'e$^4$, 
E.~Artis$^{1,5}$,
H.~Aussel$^4$, 
I.~Bartalucci$^6$, 
A.~Beelen$^7$, 
A.~Beno\^it$^8$, 
S.~Berta$^9$, 
L.~Bing$^7$, 
B.~Bolliet$^{10}$
O.~Bourrion$^1$, 
M.~Calvo$^8$, 
A.~Catalano$^1$, 
M.~De~Petris$^{11}$, 
F.-X.~D\'esert$^{12}$, 
S.~Doyle$^3$, 
E.~F.~C.~Driessen$^9$, 
G.~Ejlali$^{13}$, 
A.~Ferragamo$^{11}$,
A.~Gomez$^{14}$, 
J.~Goupy$^8$, 
C.~Hanser$^1$, 
S.~Katsioli$^{15,16}$, 
F.~K\'eruzor\'e$^{17}$, 
C.~Kramer$^9$, 
B.~Ladjelate$^{18}$, 
G.~Lagache$^7$, 
S.~Leclercq$^9$, 
J.-F.~Lestrade$^{19}$, 
J.~F.~Mac\'ias-P\'erez$^1$, 
S.~C.~Madden$^4$, 
A.~Maury$^4$, 
P.~Mauskopf$^{3,20}$, 
F.~Mayet$^1$, 
J.-B.~Melin$^{21}$
A.~Monfardini$^8$, 
M.~Mu\~noz-Echeverr\'ia$^{22}$, 
A.~Paliwal$^{11}$, 
L.~Perotto$^1$, 
G.~Pisano$^{11}$, 
E.~Pointecouteau$^{22}$, 
N.~Ponthieu$^{12}$, 
G.~W.~Pratt$^4$, 
V.~Rev\'eret$^4$, 
A.~J.~Rigby$^{23}$, 
A.~Ritacco$^{24,25}$, 
C.~Romero$^{26}$, 
H.~Roussel$^{27}$, 
F.~Ruppin$^{28}$, 
K.~Schuster$^9$, 
A.~Sievers$^{18}$, 
C.~Tucker$^3$,
G.~Yepes$^{29}$}
\address{
    \scriptsize
$^1$Univ. Grenoble Alpes, CNRS, Grenoble INP, LPSC-IN2P3, 38000 Grenoble, France \\
 $^2$Univ. C\^ote d'Azur, Observatoire de la C\^ote d'Azur, CNRS, Laboratoire Lagrange, France\\
 $^3$School of Physics and Astronomy, Cardiff University, CF24 3AA, UK \\
 $^4$Univ. Paris-Saclay, Univ. Paris Cité, CEA, CNRS, AIM, 91191, Gif-sur-Yvette, France \\
 $^5$Max Planck Institute for Extraterrestrial Physics, 85748 Garching, Germany\\
 $^6$INAF, IASF-Milano, Via A. Corti 12, 20133 Milano, Italy. \\
 $^7$Aix Marseille Univ, CNRS, CNES, LAM, Marseille, France \\
 $^8$Univ. Grenoble Alpes, CNRS, Institut N\'eel, France \\
 $^9$Institut de RadioAstronomie Millim\'etrique (IRAM), Grenoble, France \\
 $^{10}$DAMTP, Centre for Mathematical Sciences, Wilberforce Road, Cambridge CB3 
 0WA, U.K.\\
  $^{11}$Dipartimento di Fisica, Sapienza Univ. di Roma, I-00185 Roma, Italy \\
  $^{12}$Univ. Grenoble Alpes, CNRS, IPAG, 38000 Grenoble, France \\
  $^{13}$Institute for Research in Fundamental Sciences (IPM), Larak Garden, 19395-5531 Tehran, Iran \\
  $^{14}$Centro de Astrobiolog\'ia (CSIC-INTA), Torrej\'on de Ardoz, 28850 Madrid, Spain \\
  $^{15}$National Observatory of Athens, IAASARS, GR-15236, Athens, Greece \\
  $^{16}$Faculty of Physics, Univ. of Athens, GR-15784 Zografos, Athens, Greece \\
  $^{17}$High Energy Physics Division, Argonne National Laboratory, Lemont, IL 60439, USA \\
  $^{18}$Instituto de Radioastronom\'ia Milim\'etrica (IRAM), Granada, Spain \\
 $^{19}$LERMA, Observatoire de Paris, PSL Research Univ., CNRS, Sorbonne Univ., UPMC, France\\
  $^{20}$School of Earth \& Space and Department of Physics, Arizona State University, AZ 85287, USA \\
  $^{21}$Univ. Paris-Saclay, CEA, Département de Physique des Particules, 
  91191, Gif-sur-Yvette, France.\\
  $^{22}$Univ. de Toulouse, UPS-OMP, CNRS, IRAP, 31028 Toulouse, France. \\
  $^{23}$School of Physics and Astronomy, University of Leeds, Leeds LS2 9JT, UK \\ 
   $^{24}$Dipartimento di Fisica, Univ. di Roma ‘Tor Vergata’, via della Ricerca Scientifica 1, I-00133 Roma, Italy\\
   $^{25}$LPENS, ENS, PSL Research Univ., CNRS, Sorbonne Univ., Univ. de Paris, 75005 Paris, France \\
   $^{26}$Department of Physics and Astronomy, Univ. of Pennsylvania, PA 19104, USA \\
   $^{27}$Institut d'Astrophysique de Paris, CNRS (UMR7095), 75014 Paris, France \\
   $^{28}$Univ. of Lyon, UCB Lyon 1, CNRS/IN2P3, IP2I, 69622 Villeurbanne, France \\
    $^{29}$Departamento de F\'isica Te\'orica and CIAFF, Facultad de Ciencias, Univ. Aut\'onoma de Madrid, 28049 Madrid, Spain.}
\maketitle\abstracts{
In Sunyaev-Zeldovich (SZ) cluster cosmology,
two tools are needed to be able to exploit data from large scale surveys in the millimeter-wave domain. 
An accurate description of the IntraCluster Medium (ICM) pressure profile is needed along with the scaling relation connecting the SZ brightness to the mass.
With its high angular resolution and large field of view, The NIKA2 camera, operating at 150 and 260 GHz, is perfectly suited for precise cluster SZ mapping.
The SZ Large Program (LPSZ) of the NIKA2 collaboration is dedicated to the observation of a sample of 38 SZ-selected 
clusters at intermediate to high redshift and observed both in SZ and X-ray.
The current status is that all LPSZ clusters have been observed and the analysis toward the final results is ongoing. 
We present in detail how NIKA2-LPSZ will obtain a robust estimation of the SZ-Mass scaling relation and how it will be used to obtain cosmological constraints.
}
\section{Introduction}
\label{sec:intro}
Clusters of galaxies may be used as probes to constrain our cosmological model. 
They can be detected in the millimeter domain thank to the thermal Sunyaev-Zeldovich (SZ) effect {\cite{SZ_effect}}. 
It produces a distortion of the Cosmic Microwave Background (CMB) spectrum to higher frequency caused by an inverse 
Compton scattering of the CMB photons by the electrons of the IntraCluster Medium (ICM). This effect is quantified
  by the Compton parameter $y$ that is proportional to the integral of the electronic pressure $P_e(r)$ along the line of sight. 
  To constrain cosmological parameters, one method consists in building a catalog of SZ-detected clusters and count their numbers per mass and redshift bins. 
The hydrostatic mass of a cluster can be derived from a joint analysis of the Compton profile and the X-ray density profile. 
However, this is only possible for a small fraction of clusters within large millimeter surveys; from these datasets, in most cases, one only has access to $Y_{500}$, 
which is the surface-integrated Compton-$y$ parameter, 
which can be related to $M_{500}$ with a scaling relation, enabling mass calibration for all detected clusters
 \footnote{that corresponds to density 500 times the critical density of the Universe.}.
Assuming galaxy clusters are
spherical, in a hydrostatic equilibrium and their ionized gas is considered as ideal, the hydrostatic mass and $Y_{500}$ are related with a power law as follows:
\begin{equation}
    E(z)^{-\frac{2}{3}}\left(\dfrac{D_A^2 Y_{500}}{10^{-4} \mathrm{Mpc}^2}\right)=10^\alpha \left(\dfrac{(1-b)M^{\rm{HSE}}_{500}}{6\times 10^{14} \mathrm{M}_\odot}\right)^\beta
    \label{eq:YM_SR}
\end{equation}
where $b$ is the hydrostatic bias, $E(z)$ the reduced Hubble parameter and $D_A$ the angular diameter distance.
However, these assumptions do not describe the variety of clusters in the Universe.
Thus, the scaling relation is defined as a Gaussian distribution with an intrinsic dispersion: 
\begin{equation}
    P(\log (\widetilde{Y}_{{500}})|\log (\widetilde{M}_{500}))=\mathcal{N}(\alpha + \beta \log (\widetilde{M}_{500}),\sigma_{int}^2)
    \label{eq:real_SR}
\end{equation}
where $\widetilde{Y}_{500}$ corresponds to the left-hand side of equation \ref{eq:YM_SR} and $\widetilde{M}_{500}$ is equal to the term in bracket on the right-hand side.
Therefore, the SZ-Mass scaling relation is defined with three parameters, $\alpha$ the intercept, $\beta$ the slope and $\sigma_{int}$ the intrinsic dispersion. 
Throughout the years, multiple estimations of these parameters have been obtained, based on different data sets and methods {\cite{planck_YM}}$^,$ {\cite{A24}}. 
Currently, the NIKA2 collaboration is working
on a new estimation of the scaling relation to update cosmological constraints.
It will be obtained from a sample of clusters at higher redshift ($z \in [0.5,0.9]$) than previous studies, with HSE mass proxy derived from the combination of pressure profiles 
from SZ observations and density profiles from X-ray observations. 
Thanks to the high resolution at all wavelengths, we will be able to take into account the effect of cluster morphologies on the scaling relation.
NIKA2 collaboration benefits from 300 hours of guaranteed time of observation at the IRAM 30-m telescope for the SZ Large Program {\cite{LPSZ}} (LPSZ).
It associates observations in X-ray from the {\it XMM-Newton} satellite {\cite{Xm}} with SZ observations obtained with the NIKA2 camera {\cite{NIKA2-general}}, the two instruments having matching angular resolution of $6.6''$ and $17.6''$, respectively {\cite{NIKA2-performance}}.
The latter is a dual-band KID-based instrument that observes at high resolution thanks to the IRAM 30-m telescope. 
The observations at 150 GHz give access to the decrement of the SZ signal and the point source contamination can be observed at 260 GHz.
Thanks to its large field of view, clusters can be fully mapped even at intermediate to high redshift.
The LPSZ sample is composed of 38 clusters detected in the Planck \cite{Planck Cat}  and ACT \cite{ACT} surveys. 
It was constructed so that it would not be sensitive to the underlying mass distribution.
For each cluster, pressure and density profiles will be estimated. They 
 will be used to obtain their mass profiles under the assumption of the hydrostatic equilibrium. 
Similarly, the integrated $M^{\rm{HSE}}_{500}$ and $Y_{500}$ quantities will be estimated.
Ultimately, the LPSZ team will provide the sample products: maps, noise maps,
thermodynamic profiles and the integrated quantities, along with the mean pressure profile \cite{H23} and the SZ-Mass scaling relation. The latter product will be 
applied to Planck data and will take into account all systematic effects that come from the data analysis or the selection function.
\section{NIKA2-LPSZ data analysis}
Thermodynamics profiles and integrated quantities are obtained 
using the panco2 {\cite{panco2}} software with inputs of NIKA2 maps and {\it XMM-Newton} data. Then, an estimation of the scaling relation is obtained with the LIRA {\cite{LIRA}} software, 
using the following inputs: $Y_{500}$, $M_{500}$ and their covariance matrix for each cluster.
Finally, this scaling relation is applied to a large millimeter survey to obtain cosmological constraints using class-sz {\cite{class-sz}}.
As previously mentioned, the SZ-Mass scaling relation is affected by several systematics effects.
One of them is due to the cluster morphology that is one of the effect we will study thanks to the NIKA2 angular resolution. 
Other systematics effects comes from the data processing, for example the residual instrumental and atmospheric noise. 
Besides, the selection function is the systematic effect studied in this work. 
Indeed, the LPSZ clusters have been selected to fill ten 2D bins in redshift (two bins) and $Y_{500}$ values (five bins). This was done to force the sample to be homogeneous in the mass-redshift plane. 
In order to study the effect of the LPSZ selection function on the estimation of the scaling relation, 
a simulation was developped. Starting from a mass-resdshift catalog following a halo mass function with a fixed cosmology, all the clusters 
are characterized by their redshift and $M^{\rm{HSE}}_{500}$,
as well as a $Y_{500}$ value per cluster that is linked to $M^{\rm{HSE}}_{500}$ with a known scaling relation. 
Then, we applied the LPSZ selection and used LIRA to obtain an estimation of the scaling relation. These steps are repeated 
(5000 times) with the same scaling relation but a different LPSZ sample distribution in the $z-Y_{500}$ plan.
To take into account the selection function effect, the simulation has been run over a range of possible scaling relations.
\begin{figure}[t]
    \begin{minipage}[c]{.46\linewidth} 
        \center
       \includegraphics[scale=0.13]{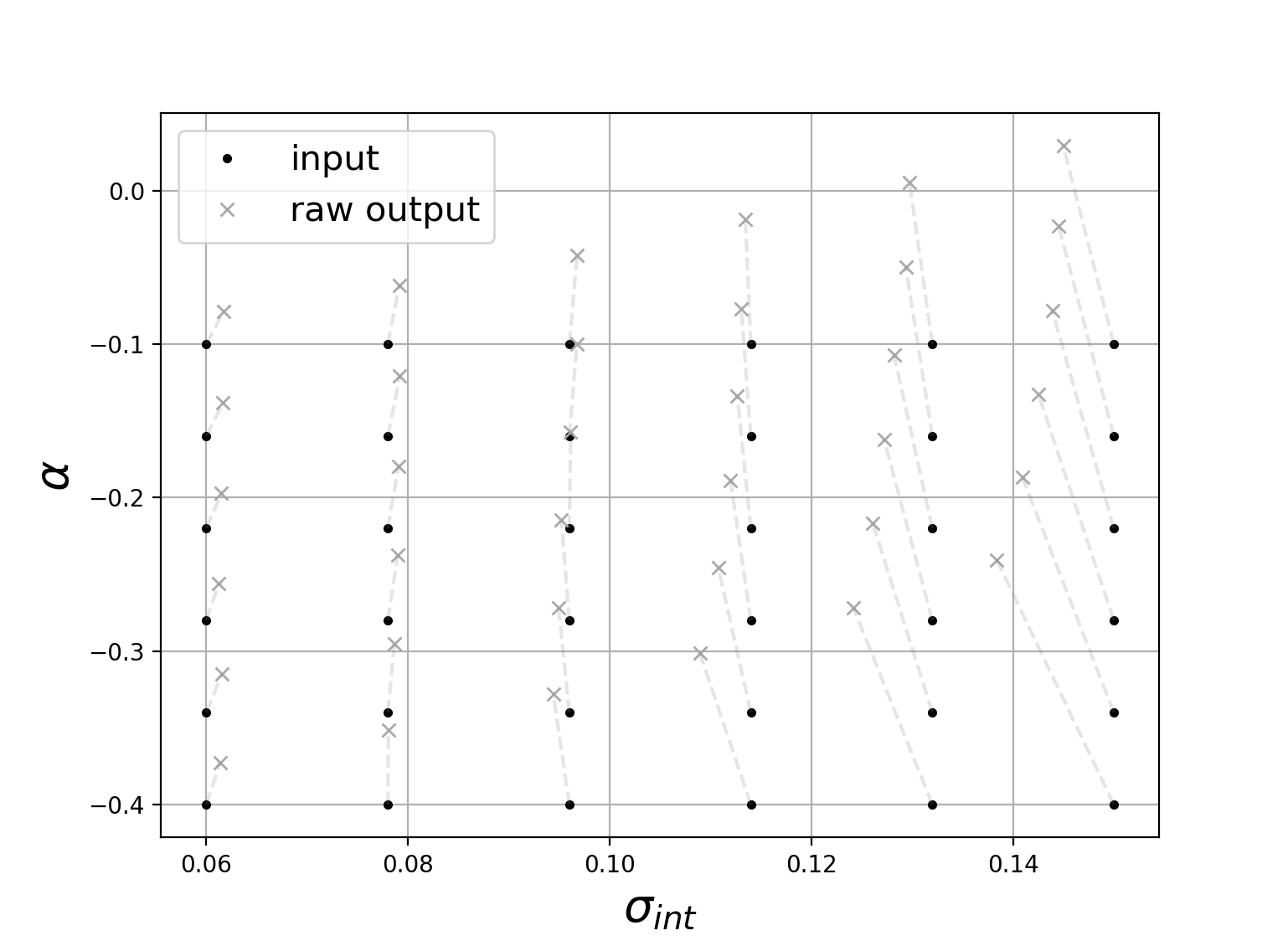} 
       \caption{\scriptsize Deviations of the recovered scaling relations (cross) from the input values (point).}
       \label{fig:deviation}
    \end{minipage} \hfill 
    \begin{minipage}[c]{.46\linewidth} 
        \center
       \includegraphics[scale=0.55]{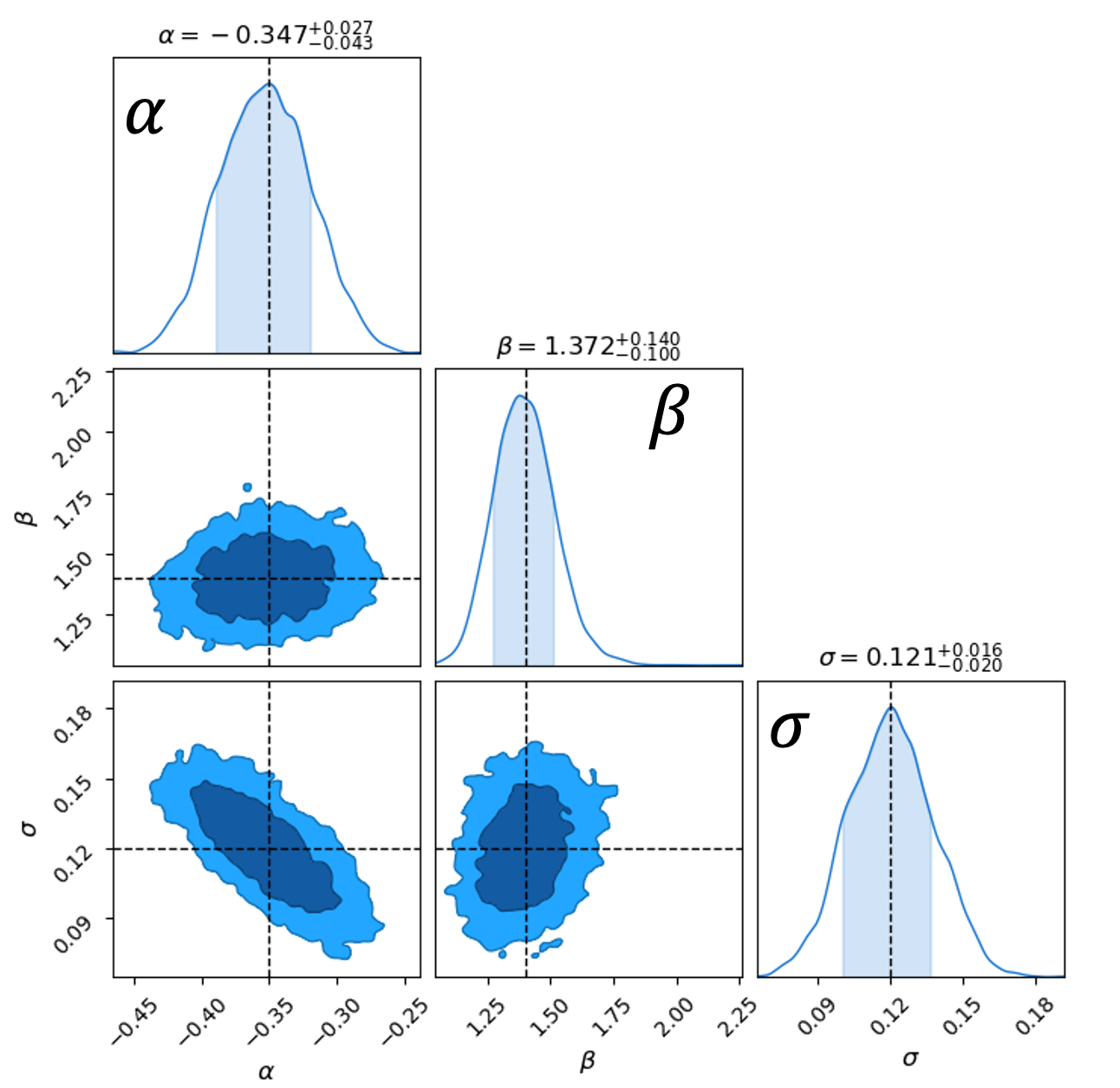} 
       \caption{\scriptsize Comparison between input (black lines) and output (blue contours) scaling relation parameters for 5000 realisations after applying the deviation factors.}
       \label{fig:results}
    \end{minipage} 
 \end{figure}
In Figure \ref{fig:deviation} we compare the input and output values of the scaling relation parameters and we observe an increasing deviation as $\sigma_{int}$ increases, which is expected \footnote{the deviation can also be observed for $\beta$ but it is not shown here for clarity.}. 
A deviation factor is defined as the ratio between the output and the input scaling relation parameters: $\alpha_{\rm{output}}/\alpha_{\rm{input}}$ for example. Then, once we have 
a raw estimation, we 3D-interpolate the deviation factors using pre-tabulated values. As shown in 
Figure \ref{fig:results}, for multiple realisations of the same input scaling relation, the input values are recovered after applying the deviation factor.
\section{Application to a cosmology sample}
\begin{figure}[t]
    \center
    \includegraphics[scale=0.18]{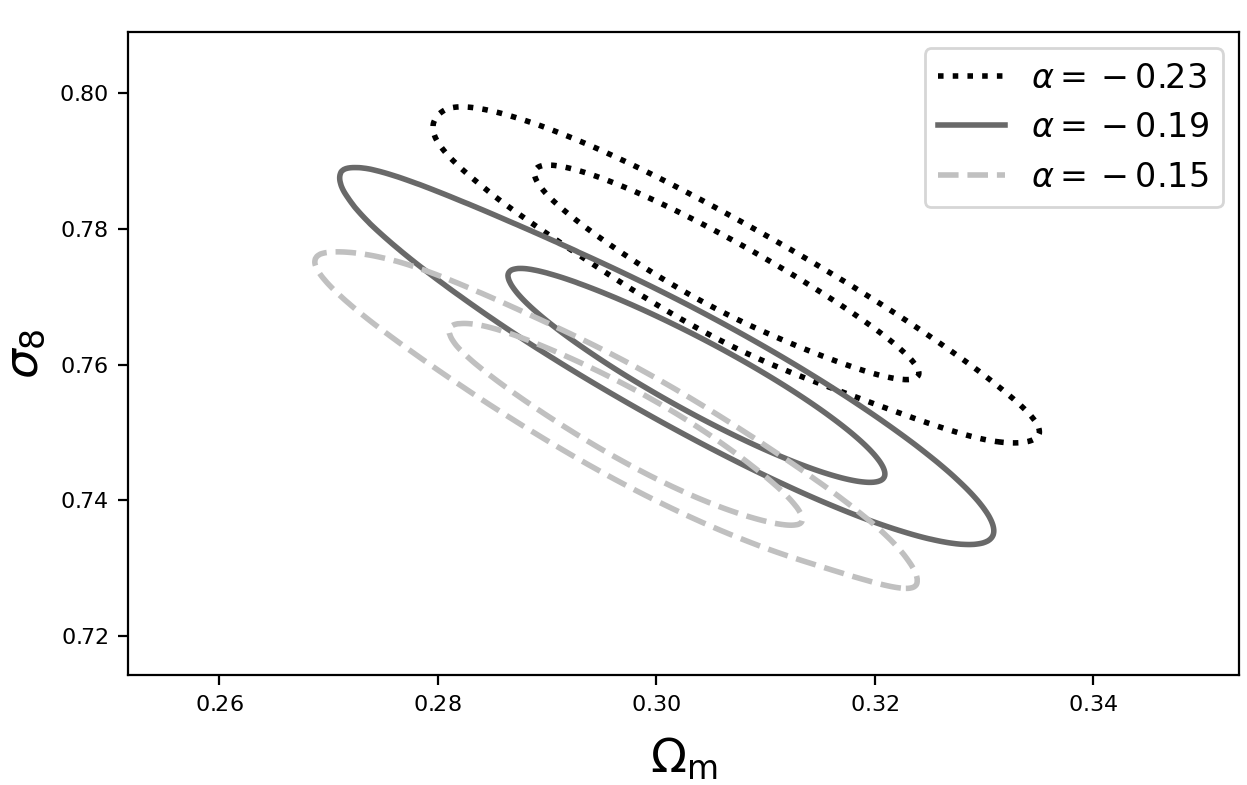}
    \caption{ \scriptsize Cosmological parameters $\Omega_m$ and $\sigma_8$ for 3 values of the intercept $\alpha$. The contours are for the 68 \% and 95 \% CL.}
    \label{fig.cosmo}
\end{figure}

The next step is to apply 
the NIKA2-LPSZ scaling relation to large millimeter surveys (Planck catalog {\cite{Planck}}) with class-sz to update cosmological constraints derived from the cluster cumber count.
For a prospective work, different input scaling relations have been tested. Figure \ref{fig.cosmo} illustrates
the dependence of the cosmological parameters with the value of $\alpha$ (the intercept). 
As $\alpha$ increases, $\Omega_m$ (the total matter density) and $\sigma_8$ (matter density fluctuations at 8 $\rm{Mpc h^{-1}}$) decrease. This can be explained by the fact that a higher $\alpha$ leads to a lower $M_{500}$ for 
a fixed $Y_{500}$ (see eq. \ref{eq:YM_SR}).
This highlights the fact that for a given set of data, the cosmological constraints on $\sigma_8$, $\Omega_m$ strongly depend
on the input scaling relation. A 2$\sigma$ deviation from the value of paper {\cite{planck_YM}} ($\alpha=-0.19\pm 0.02$) results in a disagreement in the cosmological constraints for a fixed value of the hydrostatic bias $B=1.25$
\footnote{With SZ data on its own, we cannot constrain the hydrostatic bias.}.
This is why we need, for future and present large surveys, a reliable estimation of the SZ-Mass scaling relation.

\section{Conclusion}
With a thorough study of the selection function based on simulations, we are now able to take its effects into account and retrieve the underlying scaling relation. 
We will be able to update the cosmological results from existing surveys.
The following phase is to study and identify effects due to the cluster dynamical states thanks to morphological indicators. We will also
study the possibility that the scaling relation is redshift-dependant and account for such evolution in the analysis of cluster catalogs.
Following this proceedings, more detailed papers and complete analysis of the LPSZ data presenting the final results mentionned are on their ways.

\scriptsize
\section*{Acknowledgments}
We would like to thank the IRAM staff for their support during the observation campaigns. The NIKA2 dilution cryostat has been designed and 
built at the Institut N\'eel. In particular, we acknowledge the crucial contribution of the Cryogenics Group, and in particular
Gregory Garde, Henri Rodenas, Jean-Paul Leggeri, Philippe Camus. This work has been partially funded by the Foundation Nanoscience
Grenoble and the LabEx FOCUS ANR-11-LABX-0013. This work is supported by the French National Research Agency under the contracts
"MKIDS", "NIKA" and ANR-15-CE31-0017 and in the framework of the "Investissements d’avenir” program (ANR-15-IDEX-02).
This work has benefited from the support of the European Research Council Advanced Grant ORISTARS under
the European Union's Seventh Framework Programme (Grant Agreement no. 291294).
S. K. acknowledges support provided by the Hellenic Foundation for Research and Innovation (HFRI) under
the 3rd Call for HFRI PhD Fellowships (Fellowship Number: 5357).

\end{document}